\documentclass[aps,prl,twocolumn,amsfonts,amsmath,floatfix,showpacs,superscriptaddress,footnoteintext]{revtex4}

\usepackage[final]{graphicx}
\usepackage{color}

\newcommand{\be}{\begin{equation}} \newcommand{\ee}{\end{equation}}

\begin{document}

\title{Critical Casimir Force between Inhomogeneous Boundaries}

\date{\today}

\author{Jerome Dubail}
\affiliation{IJL, CNRS \& Universit\'e de Lorraine, Boulevard des Aiguillettes
F-54506 Vand\oe{}uvre-l\`es-Nancy Cedex, France}

\author{Raoul Santachiara}
\affiliation{Laboratoire de Physique
Th\'eorique et Mod\`eles Statistiques, CNRS UMR 8626, B\^at.~100,
Universit\'e Paris-Sud, 91405 Orsay cedex, France}

\author{Thorsten Emig}
\affiliation{Laboratoire de Physique
Th\'eorique et Mod\`eles Statistiques, CNRS UMR 8626, B\^at.~100,
Universit\'e Paris-Sud, 91405 Orsay cedex, France}
\affiliation{Massachusetts Institute of Technology, MultiScale Materials Science
for Energy and Environment, Joint MIT-CNRS Laboratory (UMI 3466),
Cambridge, Massachusetts 02139, USA}
\affiliation{Massachusetts Institute of
Technology, Department of Physics, Cambridge, Massachusetts 02139, USA}

\begin{abstract}
  To study the critical Casimir force between chemically
  structured boundaries immersed in a binary mixture at its demixing
  transition, we consider a strip of Ising spins subject to
  alternating fixed spin boundary conditions. The system exhibits a
  boundary induced phase transition as function of the relative amount
  of up and down boundary spins. This transition is associated with a
  sign change of the asymptotic force and a diverging correlation length
  that sets the scale for the crossover between different universal
  force amplitudes. Using conformal field theory and a mapping to
  Majorana fermions, we obtain the universal scaling function of this
  crossover, and the force at short distances.
\end{abstract}


\maketitle

Fluctuation-induced forces are generic to all situations where
fluctuations of a medium or field are confined by boundaries. Examples
include QED Casimir forces \cite{Casimir:1948bh,Bordag:2009ve}, van
der Walls forces \cite{Parsegian:2005ly}, and thermal Casimir forces
in soft matter which are most pronounced near a critical point where
correlation lengths are large~\cite{Gennes:1978qf,Krech:1994kl}.  The
interaction is then referred to as critical Casimir force (CCF).
Analogies and differences between these variants of the common
underlying effect have been reviewed in Ref.~\cite{Gambassi:2009rw}.

Experimentally, CCFs can be observed indirectly in wetting films of
critical fluids \cite{Nightingale:1985rm}, as has been demonstrated
close to the superfluid transition of ${}^4$He \cite{Garcia:1999} and
binary liquid mixtures \cite{Garcia:2002rc}. More recently, the CCF
between colloidal particles and a planar substrate has been 
measured directly in a critical binary liquid mixture
\cite{Hertlein:2008,Soyka:2008}.  Motivated by the possibility that
the lipid mixtures composing biological membranes are poised at
criticality~\cite{Baumgart:2007,Veatch:2007}, it has been also
proposed that inhomogeneities on such membranes are subject to a
CCF~\cite{Machta:2012fu} which provides an example of a 2D
realisation.

The amplitude of the CCF is in general a universal scaling function that
is determined by the universality classes of the fluctuating medium
\cite{Diehl:1986xq}. It depends on macroscopic properties such as the
surface distance, shape and boundary conditions of the surfaces but is
independent of microscopic details of the system \cite{Krech:1994kl}.
Controlling the sign of fluctuation forces (attractive or repulsive)
is important to a myriad of applications in design and manipulation of
micron scale devices. While for QED Casimir forces a generalized
Earnshaw's theorem rules out the possibility of stable levitation (and
consequently force reversals) in most cases~\cite{Rahi:2010yg}, the
sign of the CCF depends on the boundary conditions at the confinement.
For classical binary mixtures, surfaces have a preference for one of
the two components, corresponding to fixed spin boundary conditions
($+$ or $-$) in the corresponding Ising universality class. Depending
on whether the conditions are like ($++$ or $--$) or unlike ($+-$ or
$-+$) on two surfaces, the CCF between them is attractive or
repulsive.  So-called ordinary or free spin boundary conditions are
difficult to realize experimentally but can emerge due to
renormalization of inhomogeneous conditions as we shall show below
\cite{Toldin:2013uk}. 
 
Motivated by their potential relevance to nano-scale devices,
fluctuation forces in the presence of geometrically or chemically
structured surfaces have been at the focus recently. Sign changes of
CCFs due to wedge like surface structures have been reported very
recently \cite{Bimonte:2015ph}. Competing boundary conditions can give
rise to interesting crossover effects with respect to strength and
even sign of the forces. Here we consider such a situation for the
Ising universality class in 2D. At criticality, this system can be
described by conformal field theory
(CFT)~\cite{Friedan:1984,Cardy:1989}, and CCFs are related to the
central charge of the
CFT~\cite{Cardy:1986fu,Kleban:1991ff,Kleban:1996pi}, and scaling
dimensions of boundary operators \cite{Cardy:1989xe}.

\begin{figure}[t]
\includegraphics[width= 1.\columnwidth]{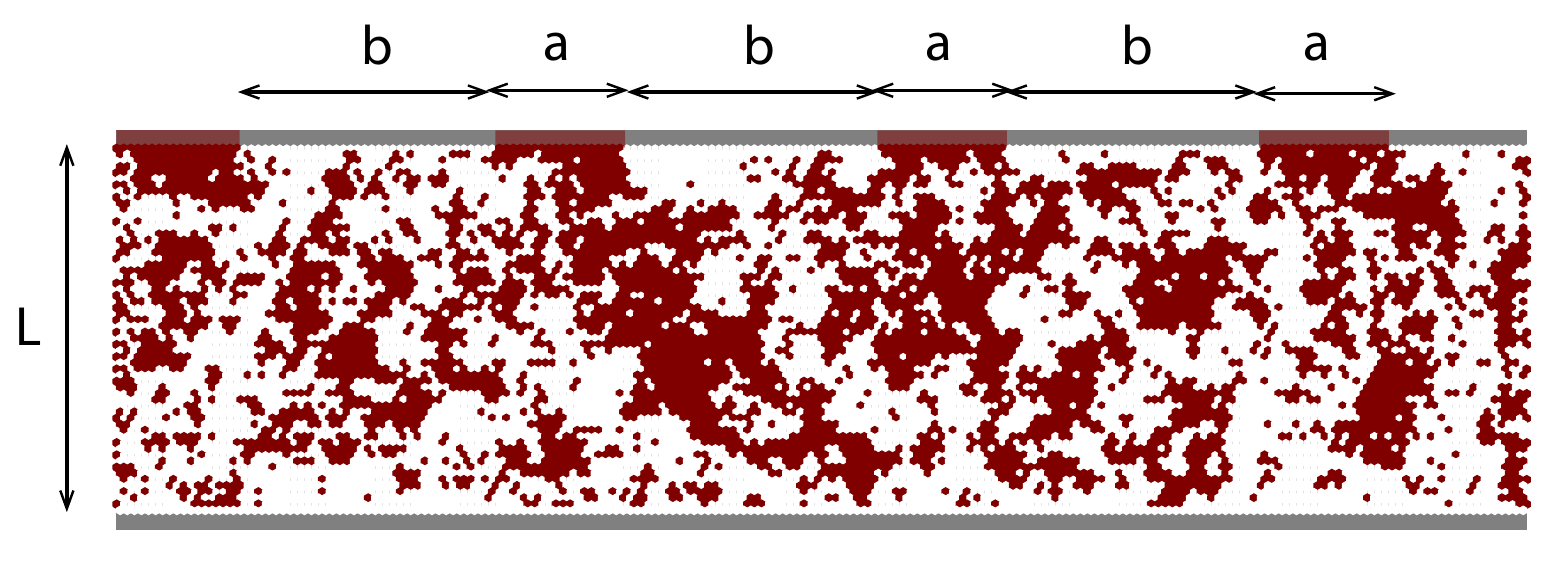}
\caption{Ising strip of width $L$ with alternating fixed spin boundary
  conditions on one side, with a typical spin configuation indicated
  by the shading.}
\label{fig:geometry}
\end{figure}

In this Letter, we show that boundary conditions which alternate
  periodically between two spin states (see Fig.~\ref{fig:geometry})
  give rise to a novel phase transition. Associated with that is a
  diverging correlation length that sets the scale for a sign change
  of the CCF on one side of the transition. We obtain the critical
  exponents and exact expressions for the universal scaling function
  of the force in the critical region.  Consider the Ising model
on an infinitely long strip of width $L$, and assume that the system is
at its critical temperature $T_c$ so that it is conformally invariant.
For homogenous, fixed spin boundary conditions $\gamma_1$,
$\gamma_2 = \pm$ on the two boundaries, the critical Casimir energy
per unit strip length, ${\cal F}$, is determined by CFT. Since $L$ is
the only finite length scale, the energy obeys a simple power law. The
amplitude is determined only by the central charge $c=1/2$ of the
Ising model and the scaling dimension $h_{\gamma_1 \gamma_2}$ of the
so-called boundary condition changing (BCC) operator from $\gamma_1$
to $\gamma_2$ (see below for details on the BCC operator)
\cite{Cardy:1989xe},
\begin{equation}
  \label{eq:1}
  {\cal F} = - \pi \left( \frac{1}{48} - h_{\gamma_1 \gamma_2} \right) \frac{1}{L}
\end{equation}
where we measure here and the following energies in units of
$k_B T_c$.  For like boundary conditions $\gamma_1=\gamma_2 = +$ or
$-$ one has $h_{++}=h_{--} = 0$ and hence an attractive force
$F = - d {\cal F}/dL$. For unlike boundary conditions
$\gamma_1 \neq \gamma_2$ one gets $h_{+-}=h_{-+}=1/2$ and hence a
repulsive force. There is one more conformally invariant boundary
condition that corresponds to free (f) spins or ordinary boundary conditions.
When combined with fixed boundary conditions, the corresponding BCC
operator has the scaling dimension $h_{f+}=h_{f-}=1/16$ which implies a
repulsive interaction in Eq.~(\ref{eq:1}). 
In the following we consider a strip with homogeneous $+$ spins
  on one boundary and alternating regions of $-$ and $+$ spins of
  length $a$ and $b$, respectively, on the other boundary, see Fig.~\ref{fig:geometry}.

If the temperature is slightly different from $T_c$, the system is in
the critical region, where the free energy density can be decomposed
into non-singular (${\cal F}_{ns}$) and singular (${\cal F}_{s}$)
contributions,
\begin{equation}
  \label{eq:2}
  {\cal F}(t,L,\tau) =   {\cal F}_{ns}(t,L,\tau)+  {\cal F}_s(t,L,\tau)
\end{equation}
that depend on the reduced temperature $t=T/T_c-1$, the width $L$, and
a scaling variable $\tau=a/b-1$ that is specific to the alternating
boundary conditions in Fig.~\ref{fig:geometry}. While the non-singular
part is an analytic function of $t$ and $\tau$, the singular part is
not. For homogeneous boundary conditions, $t$ is the only relevant
scaling variable, and in the critical region the singular part of the
free energy density is given by a universal scaling function
$\vartheta$ that depends only on $L/\xi$
\cite{Diehl:1986xq,Krech:1994kl} where
$\xi(t\to 0^\pm) = \xi_0^\pm |t|^{-\nu}$ is the bulk correlation
length with amplitude $\xi_0^\pm$ and exponent $\nu=1$ for the Ising
model. As we shall see below, the same renormalization-group (RG)
concepts apply to a novel, {\it boundary induced critical region} that
we identify for inhomogeneous boundary conditions around $a=b$.  To
focus on that region, we assume in the following that the system is at
its bulk critical point, $t=0$.  For large $L \gg a,b$ the singular
part of the free energy density can be expressed in terms of a
universal scaling function of the new correlation length
$\xi_c(\tau)=(a+b)|\tau|^{-\nu_c}$,
\begin{equation}
  \label{eq:3}
  {\cal F}_s(0,L,\tau) = \frac{1}{L} \vartheta[L/\xi_c(\tau)] \, .
\end{equation}
Below we shall determine $\vartheta$ and the exponent $\nu_c$.

BCC operators have been introduced in CFT to study systems with
discontinuous boundary conditions \cite{Cardy:1989xe}.  When inserted
on a boundary, these local operators interpolate between the different
boundary conditions on either side of the insertion point. They are
highest weight states of weight $h$ and all such states may be
realized by an appropriate pair of boundary conditions. For the
critical Ising model, the BCC operator that takes the boundary
condition from $+$ spin to $-$ spin corresponds to the chiral part of
the energy operator $\epsilon(z,\bar z)$. This can be understood
easily in the representation of the Ising model in terms of a free
Majorana fermion field $\psi(z)$ out of which the energy operator is
composed, $\epsilon(z,\bar z)=i \psi(z)\bar\psi(\bar z)$
\cite{Francesco:1997kx}: The Jordan-Wigner transformation shows that
the fermion creation and annihilation operators flip locally the spin
orientation.

Now the BCC operators permit us to relate the partition function of
the strip with alternating boundary conditions to a correlator for the
field $\psi(z)$ at positions where the boundary conditions change.  On
the upper complex plane, one has
$\langle \psi(z)\psi(z')\rangle = 1/(z-z')$ which yields (after a
conformal map) for the partition function of the strip the Pfaffian,
\begin{equation}
    \label{eq:4}
    Z = Z_0 \langle \psi(w_1) \ldots \psi(w_{2N}) \rangle = Z_0
    \text{Pf} (G) = Z_0 {\det}^{1/2} (G), 
  \end{equation}
  with $G = [\langle \psi(w_i)\psi(w_j) \rangle]_{i,j=1,\ldots,2N}$,
  where we used the Wick theorem for fermions, $w_j$ are the positions
  of the $2N$ BCC operators on the upper edge of the strip, and $Z_0$
  is the partition function of the homogenous system with $a=0$.
Due to the symmetry under translations by $a+b$, the matrix $G$ is of
block Toeplitz form, $G_{ij} = g_{i-j}$, with
\begin{equation}
  \label{eq:5}
  g_j = 
  \begin{pmatrix}
    g[j (a+b)] & g[j(a+b)-a] \\
    g[j(a+b)+a] & g[j (a+b)] 
  \end{pmatrix} \, ,
\end{equation}
where $g(w)=\pi/[2L \sinh(\pi w / (2L)]$.

The free energy density can be expressed in the thermodynamic limit as
\begin{equation}
  \label{eq:6}
  {\cal F} = -\frac{\pi}{48} \frac{1}{L} - \lim_{N\to\infty}
  \frac{1}{2N(a+b)} \log \det G \, .
\end{equation}
The Szeg{\"o}-Widom (SW) theorem for block Toeplitz matrices states that the
determinant can be expressed in terms of the matrix valued Fourier
series $\varphi(\theta) = \sum_{k=-\infty}^\infty g_k e^{ik\theta}$
as \cite{Widom:1974vn}
\begin{equation}
  \label{eq:7}
   \lim_{N\to \infty} \frac{1}{2N} \log \det G = 
\frac{1}{4\pi} \int_0^{2\pi} d\theta \log \det \varphi(\theta) 
\end{equation}
where $\det$ acts now on a $2\times 2$ matrix. It turns out that this
formula can be only applied for the case $a<b$. The reason for that is
a subtle difference between the Toeplitz matrix $G$ and the
corresponding circulant matrix $C$ that describes periodic boundary
conditions along the strip. While for $a<b$ the spectra of $G$ and
$C$  become equivalent for $N\to\infty$, for $a>b$ there exists a pair
of eigenvalues of $GC^{-1}$ that tend to zero exponentially for
$N\to\infty$, yielding an extra contribution $\delta$ that is
determined by the decay of the Fourier integral
\begin{equation}
  \label{eq:8}
   J=\frac{1}{2\pi} \int_0^{2\pi}  d\theta e^{-i j
     \theta} \left[\varphi^{-1} (\theta)\right]_{11} \sim e^{-j \delta}
   \quad \text{for} \, j\to \infty
\end{equation}
and has to be {\it subtracted} from the r.h.s. of Eq.~(\ref{eq:7}) for
$a>b$.  Here $\left[\varphi^{-1} (\theta)\right]_{11}$ denotes the
$11$-element of the $2\times 2$ matrix $\varphi^{-1} (\theta)$.
In the following we apply Eqs.~(\ref{eq:7}) and (\ref{eq:8}) to
compute the critical Casimir force in various scaling limits.

When $L\ll a,b$, the function $g(w)$ defined below Eq.~(\ref{eq:5})
can be replaced by $g(w)=(\pi/L)e^{-\pi|w|/(2L)}$ which yields the
exact determinant
\begin{equation}
  \label{eq:9}
  \det \varphi(\theta) = \frac{\cos \theta -\cosh(\pi(a-b)/(2L))}{\cos 
\theta -\cosh(\pi(a+b)/(2L))} \, .
\end{equation}
For $a<b$, the SW theorem then yields
$\frac{1}{2N}\log \det G = - (\pi a)/(2L)$. For $a>b$, this is also
the correct result as it follows from subtracting the correction
$\delta$ which follows from Eq.~(\ref{eq:8}) and $J=e^{-\pi|a-b|j/(2L)}$ as
$\delta=\pi(a-b)/(2L)$. It follows that the critical Casimir force for $L\ll
a,b$ is
\begin{equation}
  \label{eq:10}
  F = \frac{\pi}{48} \frac{23a-b}{a+b} \frac{1}{L^2} + \ldots
\end{equation}
It has an analytic amplitude that varies continously with $a/b$.  This
result is identical to an addition of the amplitudes from
Eq.~(\ref{eq:1}) for unlike and like boundary conditions, weighted by
$a/(a+b)$ and $b/(a+b)$, according to their occurrence.  Hence,
additivity holds at short distances.  This has been observed also for
a 3D Ising model in the special case of boundaries with alternating
stripes of equal width \cite{Toldin:2013uk}.

Next, we consider the case $L \gg a,b$. Using the Abel-Plana
  summation formula, it can be shown that in this
limit the elements of the matrix 
\begin{equation}
  \label{eq:11}
\varphi(\theta) =   \frac{\pi}{L}
\begin{pmatrix}
i \gamma_1(\theta) & \gamma_2(\theta) \\
-\gamma^*_2(\theta) & i \gamma_1(\theta)
\end{pmatrix}
\end{equation}
approach 
\begin{widetext}
\vspace*{-.7cm}
  \begin{equation}
  \label{eq:12}
    \gamma_j(\theta) = \frac{L}{a+b} \left\{
    \hat\gamma_j(\theta,\tau) - i^{j+1} \left[ \tanh(\theta L/(a+b))
+\tanh((\theta-2\pi) L/(a+b)) \right]\right\}
  \end{equation}
with $\hat\gamma_1(\theta,\tau)  = 1-\theta/\pi$ and
\begin{equation}
\hat\gamma_2(\theta,\tau) = \frac{1}{\pi} \left[ 
-\frac{\tau+2}{\tau+1} + e^{i \theta} (\tau+2)
{}_2F_1\left(1,\frac{1}{\tau+2},\frac{\tau+3}{\tau+2},e^{i \theta}\right)
- e^{-i \theta} \frac{\tau+2}{2\tau+3} 
{}_2F_1\left(1,\frac{2\tau+3}{\tau+2},\frac{3\tau+5}{\tau+2},e^{-i \theta}\right)
\right] \,
\end{equation}
where ${}_2F_1$ is a hypergeometric function. For $a<b$, the SW theorem
then yields the free energy density
\begin{equation}
  \label{eq:13}
 {\cal F} = -\frac{\pi}{48}\frac{1}{L}
  -\frac{1}{4\pi(a+b)} \int_0^{2\pi} \log\left\{  1 +
   \Gamma(\theta,\tau) 
\left[ \tanh\frac{\theta L}{a+b}
+\tanh\frac{(\theta-2\pi)L}{a+b} \right] \right\} d\theta \, .
\end{equation}
\end{widetext}
\vspace*{-1.0cm}
where we have subtracted a $L$-independent contribution that does not
change the force, and defined
$\Gamma(\theta,\tau)=(2\hat\gamma_1+i(\hat\gamma_2-\hat\gamma^*_2))
/(|\hat\gamma_1|^2-|\hat\gamma_2|^2)$.
In the evaluation of the integral, the correlation length
$\xi_c(\tau)$ defined above Eq.~(\ref{eq:3}) becomes important. The
integrand is exponentially localized around $\theta=0, 2\pi$ over a
small range $(a+b)/L$.  Also, it can be shown that $\Gamma$ has the
scaling property
$\lim_{\tau\to 0} \Gamma( \tau^2/\zeta,\tau) = \Gamma_0(\zeta) =
1/(1+\pi^3\zeta/32)$
for any constant $\zeta$. Hence, in the critical region of small $\tau$, or
$\xi_c(\tau) \gg a+b$, the proper scaling is obtained by setting
$\zeta=(L\tau^2)/(a+b)=L/\xi_c$ (up to a numerical coefficient),
showing that the exponent $\nu_c=2$.  In the integral, $\Gamma(\theta,\tau)$ can be replaced
by $\Gamma_0(\zeta)$ and one obtains after a simple integration the
result for the universal scaling function of Eq.~(\ref{eq:3}) when $a<b$ or $\tau<0$,
\begin{equation}
  \label{eq:14}
  \vartheta_-(\zeta) = \frac{1}{4\pi} \text{Li}_2\left(
    \frac{2}{1+\pi^3 \zeta /32}-1\right)
\end{equation}
where $\text{Li}_2(x)=\sum_{k=1}^\infty x^k/k^2$ is a polylogarithm
function. Outside the critical region $L\gg \xi_c$, one has
$\vartheta_-(\zeta\to\infty) = -\pi/48$ so that the force is fully
dominated by the boundary regions with like spins. On the contrary, for
$L \ll \xi_c$, and hence $\tau\to 0^-$, the frustration between almost
equal amounts of fixed $+$ and $-$ spins on the boundaries leads to a
renormalization to effectively {\it free} boundary conditions with
$\vartheta_-(\zeta\to 0) = \pi/24$.  For $a>b$, the correction $\delta$
yields an extra contribution $\Delta\vartheta(\zeta)$ determined by
\begin{equation}
  \label{eq:15}
  \Delta\vartheta(\zeta) \tan[\Delta\vartheta(\zeta)] = \frac{\pi^3}{32} \zeta
\end{equation}
so that the scaling function for $\tau>0$ is
$\vartheta_+(\zeta) =\vartheta_-(\zeta) +\Delta\vartheta(\zeta)$. Since
$\Delta\vartheta(\zeta\to 0)=0$, the scaling function is continuous around
$\tau=0$. For $L \gg \xi_c$, however,
$\Delta\vartheta(\zeta\to \infty)=\pi/2$ so that the system asymptotically
realizes homogenous unlike boundary conditions with
$\vartheta_+(\zeta\to \infty)=23\pi/48$.

\begin{figure}[htb]
\includegraphics[width= .8\columnwidth]{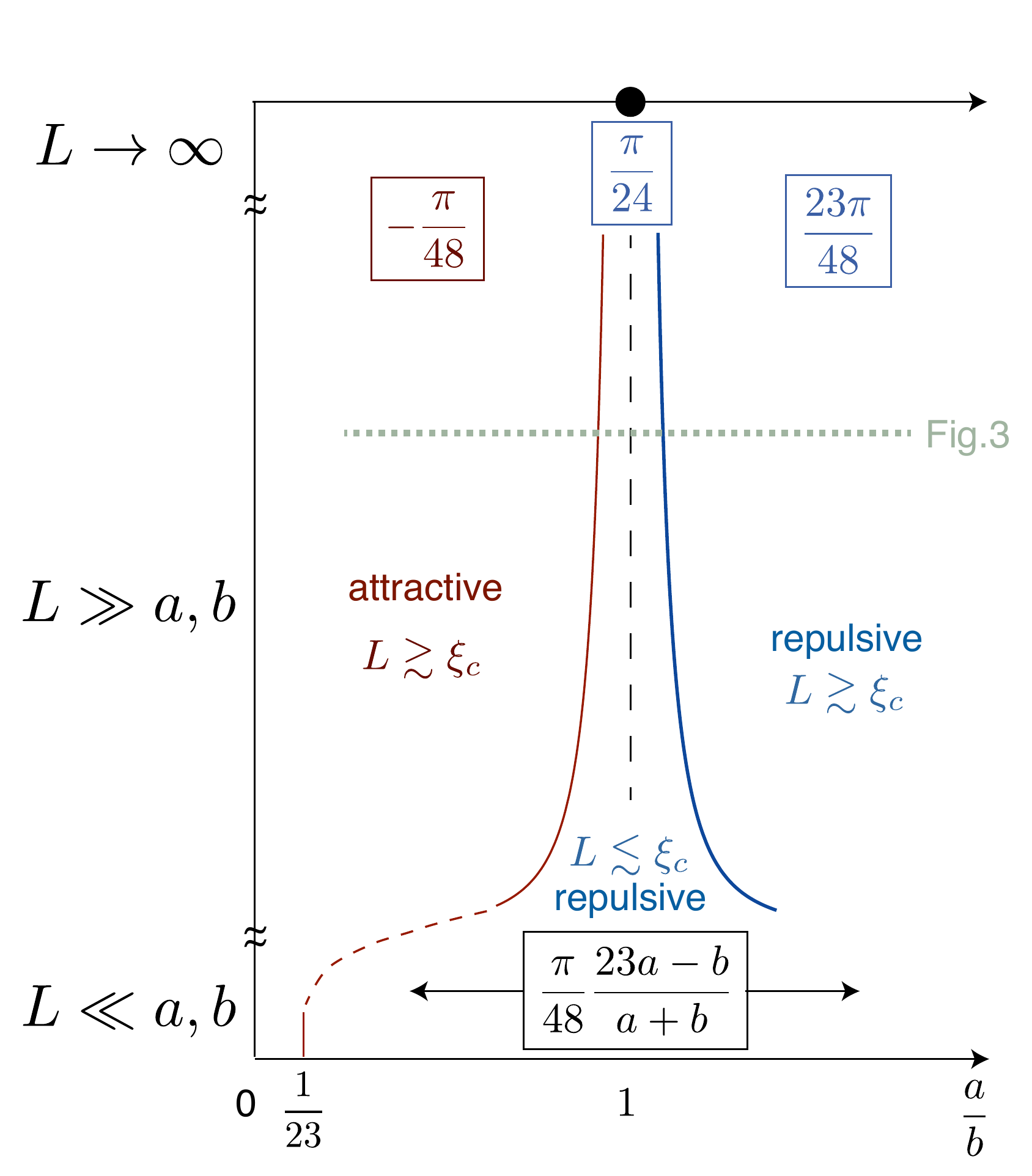}
\caption{Schematic overview of critical Casimir force amplitudes as
  function of the strip width $L$ and the ratio $a/b$. For $L\gg a,b$ the
  solid curves represent the diverging correlation length $\xi_c$. The
  horizontal dashed line indicates the cut along which the force
  amplitude is plotted in Fig.~\ref{fig:scaling_fct}. Along the red
  curve the sign of the force changes whereas the blue curve indicates
only a change between two universal (repulsive) limits.}
\label{fig:scheme}
\end{figure}

Our findings can be summarized by the scheme of Fig.~\ref{fig:scheme}.
It shows the different scaling regimes and the corresponding
asymptotic amplitudes of the Casimir force. At short distance
$L\ll a,b$ the amplitude varies continuously
across the critical point at $a=b$, with a sign change at
$b/a=23$. For $L\gg a,b$ there exist three distinct regions: around $a=b$
appears a region where $L \ll \xi_c$ where the force is repulsive and
approaches for asymptotic $L$ the universal amplitude for fixed-free
spin boundary conditions. For $a<b$, the force changes sign from
attractive to repulsive when $L$ approaches $\xi_c$, corresponding to
a stable point. For $a>b$, the force is always repulsive but the
amplitude crosses over from $\pi/24$ to $23\pi/48$ under an increase
of $L$ beyond $\xi_c$. 

\begin{figure}[tb]
\includegraphics[width= 1.\columnwidth]{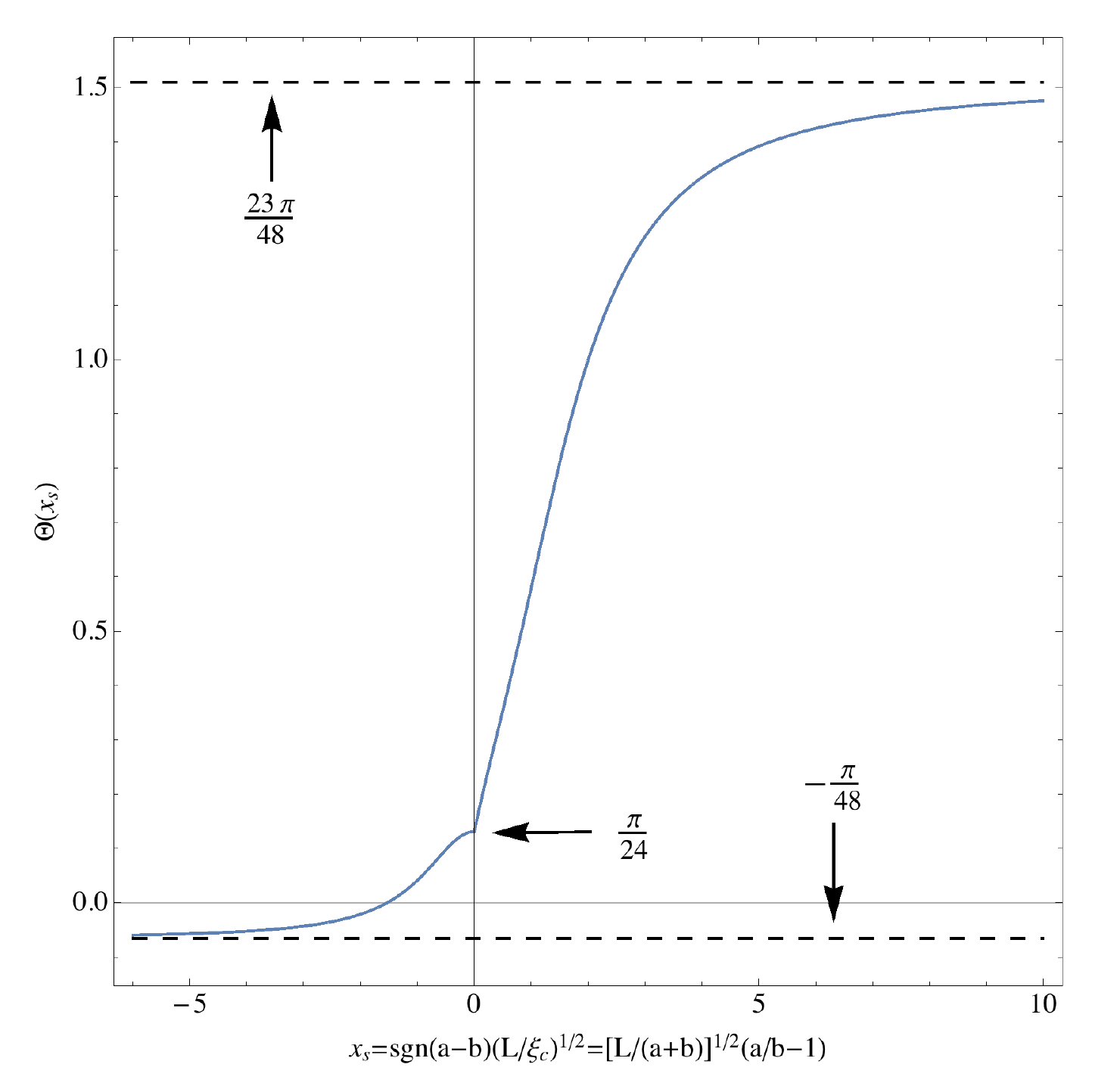}
\caption{Universal scaling function $\Theta(x_s)$ for the critical
  Casimir force as function of the scaling variable $x_s\sim a-b$.}
\label{fig:scaling_fct}
\end{figure}

The dependence of the force $F$ on $|a-b|$ at fixed $L\gg a,b$ (see
dashed horizontal line in Fig.~\ref{fig:scheme}) is determined by
$F = -\partial {\cal F}/\partial L = \Theta(x_s) L^{-2}$ with a
universal scaling function $\Theta$ of the scaling variable $x_s$
that is defined on both sides of the critical point by
 $x_s=\text{sign}(\tau)(L/\xi_c)^{1/2}\sim a-b$. This function is shown in
Fig.~\ref{fig:scaling_fct} where we used the results for
$\vartheta_\pm(L/\xi_c)$ of Eqs.~(\ref{eq:14}), (\ref{eq:15}).  In the
critical region $|x_s| \ll 1$, one has the expansions
\begin{equation}
  \label{eq:16}
  \Theta(x_s)  = \left\{
\begin{array}{ll}
\frac{\pi}{24} - \frac{\pi^2}{64} x_s^2 + \ldots & \text{for } x_s < 0 \\[1em]
\frac{\pi}{24} + \frac{\pi^{3/2}}{8\sqrt{2}} x_s + \ldots & \text{for } x_s > 0 
\end{array}\right. \, ,
\end{equation}
whereas for $L$ outside the critical region, $|x_s| \gg 1$,
\begin{equation}
  \label{eq:17}
  \Theta(x_s)  = \left\{ \!
\begin{array}{ll}
-\frac{\pi}{48} + \frac{32\log 2}{\pi^4} \frac{1}{x_s^{2}} + \ldots & 
\text{for } x_s < 0 \\[1em]
\frac{23\pi}{48} - \frac{32(\pi^2-\log 2)}{\pi^4} \frac{1}{x_s^{2}} + 
\ldots & \text{for } x_s > 0 
\end{array}\right.  .
\end{equation}
We see that $\Theta(x_s)$ is not analytic around $x_s=0$ and hence
constitutes the singular part of the free energy density, see
Eq.~(\ref{eq:2}). This resembles the singular nature of scaling
functions describing the bulk transition at $T=T_c$.

Our results show the existence of a novel phase transition for the
critical Casimir force in the 2D Ising model that is induced by
inhomogeneous boundary conditions with a varying ratio of up and down
spins.  We obtained exact expressions for the universal scaling
function of the force. Due to the observed renormalization of boundary
conditions, in binary mixtures, ordinary (free spin) boundary
conditions can be realized experimentally and ``switched'' on and off
by varying the distance $L$, or an inhomogeneous surface field.  The
crossover between different universal amplitudes leads to a stable
equilibrium point at $L\simeq \xi_c$ for $1/23<a/b<1$.  The emergence
of the novel phase transition at $a=b$ is related to the relevance of
a surface magnetic field $\sim \tau$ at a surface with free spin
boundary conditions. This can be seen from the decay of the spin
correlations along a single surface,
$\langle \sigma_x \sigma_{x'}\rangle \sim |x-x'|^{-\eta_\|}$ with
$\eta_{\|} = 1$ for free boundary conditions
\cite{Cardy:1984nr,Diehl:1986xq}. Since the surface field contributes
an energy $\sim \tau \int dx \sigma_x$, the scaling dimension
$y_\tau = 1/\nu_c = 1-\eta_{\|}/2$ which is identical to our findings
above.  It is interesting to explore these concepts in general spatial
dimensions for Ising and XY models, and tri-critical points which have
an even richer spectrum of possible boundary conditions.

We thank M.~Kardar for many fruitful discussions.

\bibliographystyle{apsrev}
\bibliography{2D_bib}

\end{document}